\documentclass[aps, prl, reprint ]{revtex4-1}

\usepackage{bm}
\usepackage{epsfig}
\usepackage{amsmath}
\usepackage{latexsym}
\usepackage{slashed} 
\usepackage{mathrsfs}
\usepackage{graphicx}
\usepackage{color}

\begin{document}

\title{{\large Phase of Quark Condensate and Topological Current in QCD  }}

\author{Chi Xiong}
\email[]{xiongchi@ntu.edu.sg}
\affiliation{Institute of Advanced Studies, \\Nanyang Technological University, Singapore 639673 }


\begin{abstract}
We propose a new topological charge term in QCD based on flux-tube models.  It couples a superflow of the phase of the quark condensate to the Chern-Simons current. The usual $\theta$-parameter is replaced by the phase of the quark condensate, which becomes nontrivial due to the existence of topological defects such as vortices. This new formulation can address the U(1) problem in a similar way as in the $\theta$-world and may avoid the strong CP problem as the new topological charge term is a derivative coupling. Some phenomenological applications like chiral magnetic effects and possible connections to the Josephson effect are discussed.

\end{abstract}

\maketitle               

Quantum Chromodynamics (QCD), the theory of strong interaction for quarks and gluons, has a very complicated vacuum with non-trivial topology. Besides the important roles it plays in studying  fundamental questions in QCD, such as confinement and chiral symmetry breaking, the topological charge has direct applications in hadron structures, for instance, the U(1) problem or the $\eta$-mass problem \cite{Weinberg:1975,'tHooft:1976, Witten:1979, Veneziano:1979, VV:1980} --- The $U_A (1)$ symmetry of the QCD Lagrangian at the chiral limit was expected to be realized in the Goldstone mode and an additional massless pseudoscalar should have existed, similar to the pions which are the Goldstone bosons from the spontaneously broken $SU_R (N_f) \times SU_L (N_f)$ chiral symmetry. However, no such flavor-singlet pseudoscalar has been observed experimentally. The candidate $\eta$-meson, with the correct quantum number, has mass violating the theoretical bound $m_{\textrm{\tiny{U(1)}}} \leq \sqrt{3} m_{\pi} $ \cite{Weinberg:1975}. The solutions to this problem include the one with instantons \cite{'tHooft:1976}, and those with the large-$N_c$ expansion of QCD \cite{Witten:1979, Veneziano:1979, VV:1980}. It has been realized that the solution to the U(1) problem should be connected to the axial anomaly. These solutions need a CP violating $\theta$-term to describe the axial anomaly. The experimental bound is $\theta < 10^{-19}$. It is interesting albeit puzzling that such a small value parameter is needed in the strong interaction and this is the strong CP problem, for which the axion particle and the Pecci-Quinn mechanism were introduced \cite{Cheng}.

In this article we propose an alternative solution to the U(1) problem. It is different from the previous approaches in that the $\theta$-parameter is not introduced in our solution, instead we have a phase angle $\alpha(\vec{x},t)$ from the quark condensate $\langle \bar{q} q\rangle$, from which a superflow $J_{\mu} \sim \partial_{\mu} \alpha $ is built to couple to the Chern-Simons current $K^{\mu}$ as
\begin{eqnarray} \label{cc}
S_{\textrm{\tiny{eff}}} &\sim & \int d^4x\,  \partial_{\mu} \alpha \,K^{\mu} \\
&=& -\int d^4x\, \alpha F \tilde{F} - \int_{{\Sigma}^a} \alpha K^{\mu} d\sigma_{\mu}^a 
\end{eqnarray}
where the familiar topological charge term $F \tilde{F}$ appears after integration by part. $\{\Sigma^a\}$ are surfaces enclosing all topological defects in the quark condensate and the infinity. 
The topological term (\ref{cc}) describes an effective interaction between the quark condensate and the gluon fields. Being a derivative coupling, it is invariant under any constant shift $\alpha \rightarrow\alpha + \alpha_0$, hence, the background value of $\alpha$, if exists, has no observable effect. This seems to be promising in addressing the strong CP problem, but let us first consider the U(1) problem in some detail. 

Our solution to the U(1) problem is based on the well-known dual-superconductor description of the QCD vacuum and QCD chromoelectric flux-tube model, in which the color electric field lines between some color sources, such as a quark and antiquark pair, are squeezed into a thin flux tube and results in a linear potential between the color sources. This linear potential is supposed to be responsible for color confinement. If these field lines are strong enough, the chiral symmetry could be restored inside the flux-tube. In \cite{Suganuma} it has been calculated that when the color electric field exceeds a critical value $E_c \approx 4$ GeV/fm, inside the flux-tube there will be a chiral symmetric phase with vanishing quark condensate $\langle \bar{q} q\rangle = 0$. This is easy to understand since a sufficiently large color electric field can pull apart the  $\bar{q}q$ pairs and destroys the quark condensate. 

The spacetime-dependence of the quark condensate changes the field configuration significantly. 
The usual chiral field should be parametrized as
\begin{equation} \label{Sigma}
U = \langle \bar{q} q\rangle \exp{[i \alpha + i \sqrt{2}/F_{\pi} ( \pi_a \lambda_a + \eta_0 /\sqrt{N_f})}]
\end{equation}
where $\lambda_a$ are the $SU(N_f)$ generators and $\eta_0$ is the $SU(N_f)$ singlet field. Eqt (\ref{Sigma}) shows that even we set $\pi = \eta_0 = 0$, $U$ still can describe vortex-like configurations of the quark condensate, e.g. a vortex with a winding number $m$ when taking $\alpha = \pm m \phi, m=1, 2, \cdots$ and $\langle \bar{q} q\rangle = f(\rho)$ with ($\rho, \phi$) being the polar coordinates on the plane transverse to the color flux-tube. For simplicity we will only consider  vortices with winding number $m=1$. The spacetime dependence of the quark condensate suggests that the boundary conditions for the function $f(\rho)$ should be $f(0)=0, f(\infty) = v$ ($v$ is a non-vanishing constant) 
\begin{equation}
\left. U \right\vert_{\pi =\eta = 0} = f(\rho) e^{i \alpha} 
\end{equation}
Now if we consider quark fields interacting with this vortex background and an Abelianized gluon potential $Z_{\mu}$, we find that it can be described by Callan and Harvey's axion string model \cite{Callan-Harvey, XC}
\begin{equation} \label{eff}
\mathcal{L}_{\textrm{\tiny{C-H}}} = \bar{q} \big[ i \gamma^{\mu}(\partial_{\mu}  - i g Z_{\mu}) - f(\rho) e^{ i \alpha \gamma^5}  + \cdots \big] q  
\end{equation}
where the ellipsis includes meson channels. It is a quite lengthy procedure to derive (\ref{eff}) from the original QCD Lagrangian. One possible way is to obtain first a {\it gauged} and non-local Nambu-Jano-Lasinio effective action, from which a chiral Lagrangian then is derived with the usual bosonization techniques. The Abelianized gauge potential $Z_{\mu}$ is built for a gauge-invariant description of the flux-tube \cite{XC}. Here we skip those steps and simply start with the action (\ref{eff}).    
It is easy to see that chiral zero modes of quarks are localized in the vortex, since they have an exponential profile $\psi_L = \chi_{L} \, \exp \big[- \int_0^{\rho} f(\rho') d\rho' \big]$ where $\chi_{L}$ is a two-dimensional spinor. The chiral zero modes are also coupled to the Abelianized gluon potential $Z_{\mu}$, therefore a gauge anomaly appears in the vortex $\mathcal{D}^k J_k = \frac{1}{2 \pi} \epsilon^{ij} \partial_i Z_j, ~(i,j,k = 0,3).$ This can be cancelled by an effective action \cite{Callan-Harvey}
\begin{equation} \label{cseff}
S_{\textrm{\tiny{C-S}}} = -\frac{ g^2 N_f}{16 \pi^2}\int d^4 x \, \partial_{\mu} \phi \, K^{\mu}
\end{equation}
which is exactly the new topological coupling (\ref{cc}) that we proposed ($\alpha =  \phi$). This cancellation happens  because the massive quark modes which live off the vortex mediate an effective interaction between the quark condensate and the gluon field, which induces a vacuum current $J^{\textrm{\tiny{ind}}}_{\mu} =  \frac{g^2 N_f}{8 \pi^2} \epsilon_{\mu\nu\rho\tau} Z^{\rho\tau} \partial^{\nu} \phi$ \cite{Callan-Harvey}. Converting it to an effective action we have
\begin{equation}
S_{\textrm{\tiny{C-S}}} = \int d^4 x \, J^{\textrm{\tiny{ind}}}_{\mu}\,Z^{\mu} = -\frac{g^2 N_f}{16 \pi^2}\int d^4 x \, \partial_{\mu} \phi \, K^{\mu}.
\end{equation}
A few remarks are in order: I. The effective coupling between the Chern-Simons current and the superflow (\ref{cseff}) emerges at quantum level, because the vortex in the quark condensate arises at quantum level; II. In contrary to the usual gauge-invariant $\theta F \tilde{F}$ term, (\ref{cc}) is {\it not} gauge-invariant. This is, however, the key of our new formulation ---  we need the gauge variance of (\ref{cc}) to cancel out the localized gauge anomaly in the topological defect, in our case the vortex, but there might be more general ones (see below and Fig.1 for other possible configurations); III. The phase of the quark condensate consists of contributions from its topological defects, $\alpha$,  and the pseudoscalar mesons as shown in (\ref{Sigma}). The pion-like Goldstone modes are small oscillations of the condensate in the direction of symmetry transformation which do not cost energy. The phase of the condensate $\alpha$ plays an important role since it couples to the Chern-Simons current  $K^{\mu}$ via (\ref{cc}), and the ghost pole in the correlator $\langle K_{\mu}K_{\nu} \rangle$ \cite{Luscher78} mixes with the singlet pseudoscalar meson, i.e. the Goldstone prototype of $\eta'$, to give the massive $\eta'$ meson. 
It is convenient to use the chiral dynamics in the large-N limit and to see how our approach can solve the U(1) problem similar to the approach in \cite{VV:1980}.  The effective Lagrangian in the large-N limit 
\begin{equation}
\mathcal{L}_{N} = - \theta q(x) + \frac{i}{2} q(x) \textrm{Tr}(\log U - \log U^{\dagger}) + \frac{N}{a F^2_{\pi}} q^2(x) + \cdots
\end{equation}
where $q(x)$ is the topological charge density and the ellipsis include other non-topological terms. It is straightforward to replace the $\theta$ parameter by the phase of the quark condensate $\alpha$, which can not be rotated away due to the topological obstacles, then eliminate the field $q(x)$ through its equation of motion and continue the rest of calculations to get the correct mass term for the $\eta'$-meson as in \cite{VV:1980}.  

The mathematical structure of our formulation is the descent equation \cite{DE}, which relates the chiral anomaly in (2n+2)-dimensions to the gauge anomaly in 2n-dimensions. 
\begin{equation}
\delta\int_{M} d\phi \wedge \mathcal{K}^0_{3} = \int_{M} d\phi \wedge d \mathcal{K}^1_{2} = - \int_{M} d^2\phi \wedge  \mathcal{K}^1_{2} = - \int_{\Sigma} \mathcal{K}^1_{2},
\end{equation}
where $ \mathcal{K}^0_{3}$ is the Chern-Simons 3-form and connected to the lower dimensional gauge anomaly $\mathcal{K}^1_{2} $ through the descent equation $\delta \mathcal{K}^{i-1}_{2n-i} = d\, \mathcal{K}^{i}_{2n-i-1} $ \cite{DE}. This can be applied to the codimension-two topological defects, such as the examples shown in Fig.1. For codimension-one topological defect like domain walls, other anomalies like parity anomaly should be taken into account \cite{Callan-Harvey}, and a similar anomaly cancellation mechanism should apply as well \cite{TX1, TX2}.

Our formulation is based on the localization of chiral zero modes of fermions. Actually the localization mechanism does not matter. We have studied the flux-tube models where the vortices in the quark condensate is induced by the strong color electric fields. How about other gauge configurations? The instanton liquid model \cite{Schafer:1996} might provide another type of localization scenario, where 't hooft has shown that chiral zero modes of fermions are located in the instanton \cite{'tHooft:1976} with a localizing profile $\lambda^{3/2}/[(x-a)^2+\lambda^2 ]^{3/2}$ ($\lambda$ and $a$ are the instanton parameters), in comparison with the exponential profile of the chiral zero modes trapped in the vortex. In the instanton liquid model, instantons interact through the exchange of quarks, so the quark zero modes can become delocalized and propagate by ``hopping" on (anti)instantons, hence, lead to some collective motion of quark condensate \cite{Schafer:1996}. Therefore QCD flux-tubes and instanton liquid have something in common --- chiral zero modes of quarks are localized in some topological defects or their ensembles. However, flux-tube models seem to have the advantage of directly addressing the confinement problem, so we formulate our model based on the flux-tubes.

\begin{figure}[!h]
\begin{center}
    \includegraphics[height=75mm, width=58mm]{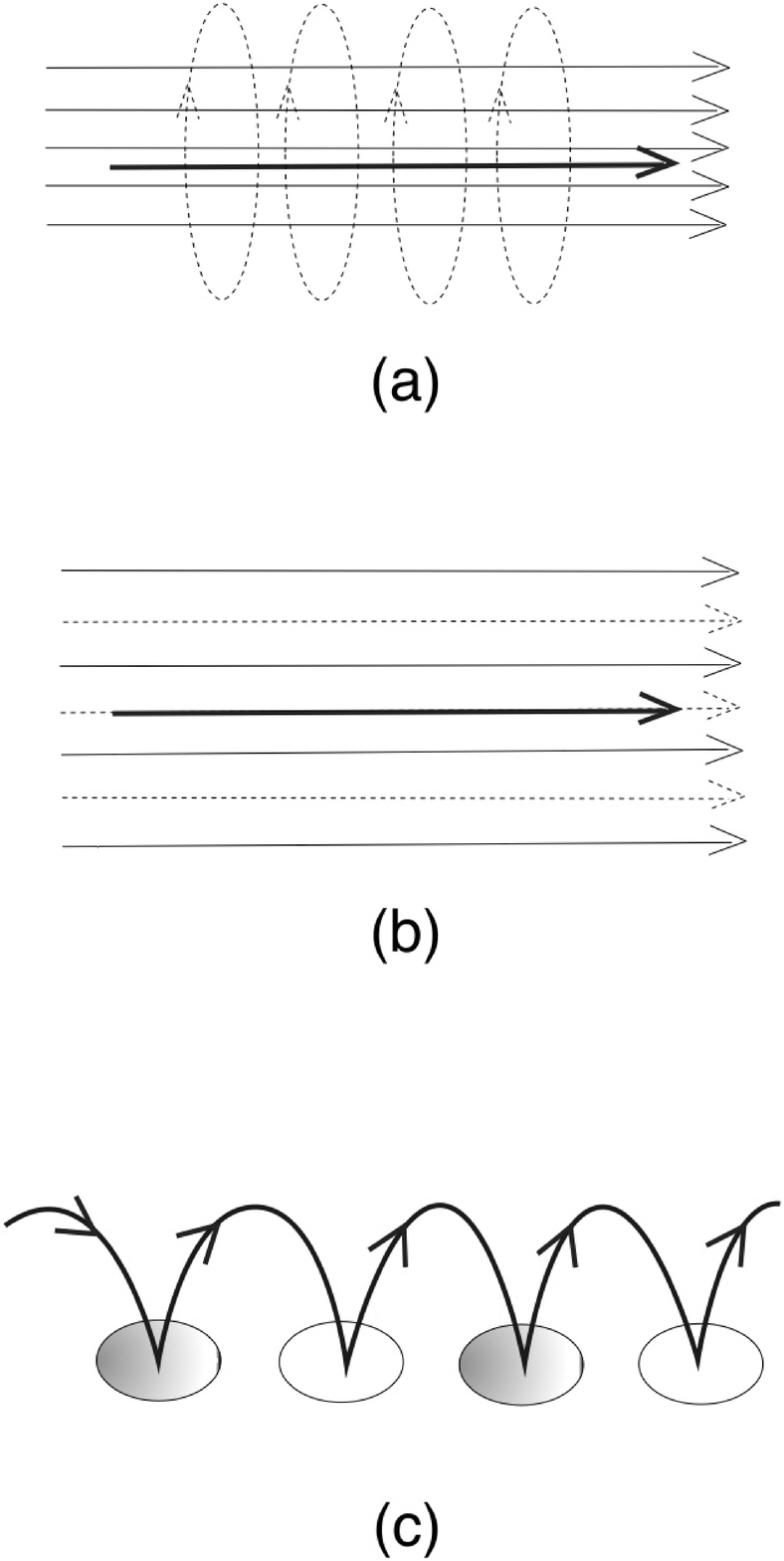}
    \caption{A sketch of different gauge configurations and localizations of chiral zero modes of quarks: (a) color electric flux-tube; (b) collinear color- electric and magnetic flux-tube; (c) a flowline in the instanton liquid where quarks travel by hopping on the instantons and the anti-instantons. The solid thick lines represent the motion of the quarks. The thin solid lines and dotted lines represent color electric and magnetic fields respectively. The grey circles and white circles are instantons and anti-instantons respectively. }
\end{center}
\label{fluxtubeInflow}
\end{figure}

The topological defects of the quark condensate might also be induced by other factors, for example, rotations and collisions. It is well known that a vortex lattice will arise in a rotating superfluid, like liquid Helium II and Bose-Einstein condensates, when the angular velocity is above some critical velocity. For ultra-relativistic collision of heavy nuclei, net color charges develop and flux tubes, with both color electric and magnetic field ($\vec{E} \cdot \vec{B} \neq 0)$, connect the color charges. For example, in \cite{Lappi06} longitudinal color electric and magnetic fields are produced after high energy hadronic collisions. Such gauge configurations carry topological charges and lead to net chiral or antichiral quark zero modes in the flux tubes. An interesting phenomenon is that they can separate charge in a background (electromagnetic) magnetic field, and consequently, an electromagnetic current is generated along the magnetic field. This is called the chiral magnetic effect \cite{Kharzeev07, Fukushima08}. It is easy to see that we should have an extra effective coupling 
\begin{equation} \label{mcs}
\Delta\mathcal{L}_{\textrm{\tiny{MCS}}} \sim -\frac{e^2}{8 \pi^2} \, \partial_{\mu} \alpha \, K^{\mu}_{\textrm{\tiny{MCS}}}
\end{equation}
where $ K^{\mu}_{\textrm{\tiny{MCS}}} =  \epsilon^{\mu\nu\rho\tau} A_{\nu} F_{\rho\tau} $ is the Maxwell-Chern-Simons current. This leads to a modified Maxwell equation
\begin{equation}
\partial_{\mu} F^{\mu\nu} = J^{\nu} - \frac{e^2}{2 \pi^2} \, \partial_{\mu} \alpha \, \tilde{F}^{\mu\nu}
\end{equation}
and the induced current $\vec{J} \sim -\dot{\alpha} \vec{B}^{\textrm{\tiny{M}}}$ agrees with \cite{Kharzeev07, Fukushima08}.

We close by comparing our theory with the axion model and the Pecci-Quinn mechanism \cite{Cheng}. The formulations seem to be similar, however, we do not introduce any new particles and what we need is some topological defect of the quark condensate which leads to a nontrivial phase distribution. The shift symmetry is taken care of by the derivative coupling, and the gauge invariance of the whole theory is ensured by the descent equation. It is also interesting to notice that if there are a cluster of quark condensates separated by some junctions, each of them has a different phase, resembling the Josephson effect in superconductors with Josephson tunnelling junctions. The Josephson critical current $\sim \cos(\Delta \theta_{ij})$ may correspond to some axion-potential $ \sim  \cos(\Delta \alpha_{ij})$ in QCD  where $\Delta \theta_{ij}$ and $\Delta \alpha_{ij}$ are their phase differences, respectively.  We leave this topic to future investigations.
 
The author thank Hank Thacker, Peter Arnold, Kerson Huang and Peter Minkowski for valuable discussion. This work is supported by the research funds from the Institute of Advanced Studies, Nanyang Technological University, Singapore.

\end{document}